# Scalable Metagrating for Efficient Ultrasonic Focusing

Y. K. Chiang,[1]* L. Quan,[2] Y. Peng,[3] S. Sepehrirahnama,[4] S. Oberst,[4] A. Alù,[2,3,5] D. Powell[1]*

[1]School of Engineering and Information Technology
University of New South Wales, Canberra, Australia
[2]Department of Electrical and Computer Engineering
The University of Texas at Austin, Austin, Texas, USA
[3]Photonics Initiative, Advanced Science, Research Center
City University of New York, USA
[4]Centre for Audio, Acoustics and Vibration
University of Technology Sydney, Australia
[5]Physics Program, Graduate Center
City University of New York, USA

*Corresponding author. E-mail: y.chiang@adfa.edu.au; david.powell@adfa.edu.au.

**Abstract**
Acoustic focusing plays a pivotal role in a wide variety of applications ranging from medical science to nondestructive testing. Previous works have shown that acoustic metagratings can overcome the inherent efficiency limitations of gradient metasurfaces in beam steering. In this work, we propose a new design principle for acoustic metalenses, based on metagratings, to achieve efficient ultrasonic focusing. We achieve beam focusing by locally controlling the excitation of a single diffraction order with the use of adiabatically varying metagratings over the lens aperture. A set of metagratings is optimized by a semi-analytical approach using a genetic algorithm, enabling efficient anomalous reflection for a wide range of reflection angles. Numerical results reveal that our metalens can effectively focus impinging ultrasonic waves to a focal point of FWHM = $0.364\lambda$. The focusing performance of the metalens is demonstrated experimentally, validating our proposed approach.

**Introduction**
Acoustic beam focusing is prominent in diverse applications, such as in medical imaging and therapy (*1-3*), nondestructive inspection of cracks in materials (*4-6*) and energy harvesting (*7, 8*). The convergence of the impinging acoustic energy in a tight focal spot is attributed to the wavefront manipulation capability of acoustic lenses. Acoustic phased-arrays are conventionally used to control the wavefront and to perform focusing (*9-11*), through controlling the amplitude and phase of individual transducers. However, the ultrasonic transducer size is generally greater than the wavelength of operation, limiting the spatial resolution of the generated field. Moreover, this conventional approach requires rather bulky phased-arrays and relatively large-scale driving electronics, leading to high fabrication cost and complex implementation.

In recent years, acoustic gradient metasurfaces with exotic beam steering properties have attracted significant attention. Phase gradient metasurfaces are formed by an array of closely-packed structures with varying geometric parameters. They enable wavefront steering by carefully adjusting the spatial phase gradient along the metasurface based on the generalized Snell's law (*12*, *13*). Such efficient beam steering facilitates the development of metasurface-based acoustic lenses, which therefore remove the requirement of conventional phased arrays. The phase gradient can be implemented by using space-coiling or helical structures, which accumulate large phase shift by controlling the propagation path (*14-20*), or Helmholtz-resonators, which modulate the phase delay based on their resonant features (*21-23*). The design of gradient metasurface lenses has also been extended to the ultrasonic range by using a set of space-coiling metamaterials with encoded prerequisite phase delays to focus acoustic beams and achieve acoustic levitation (*24*). However, the efficiency and resolution of this gradient metasurface approach greatly depends on the level of discretization. Full control of the phase change along the metasurface requires a very dense array of fine elements. Furthermore, the complex and fine internal geometries of these structures are hard to scale down to higher frequencies, and could induce significant losses in the acoustic thermal and viscous boundary layers near the walls of the structure (*25*, *26*). This typical gradient metasurface approach suffers from higher order effects due to the impedance mismatch between incident and scattered waves for large steering angles (*13*, *27*), which results in low focusing efficiency for lenses with large numerical aperture (*28*).

To address the limitations of gradient metasurfaces, an alternative class of metamaterials, known as metagratings, has recently been proposed (*29*, *30*). Metagratings are composed of periodic arrays of asymmetric bianisotropic scatterers. Their working principle of wavefront manipulation relies on the engineering of scattering into different diffraction orders. Previous works showed that acoustic metagratings enable almost 100% efficiency of anomalous reflection (or refraction) for a wide range of steering angles by using specifically engineered bianisotropic elements (*31-34*). Metagratings are advanced periodic structures designed in accordance with the grating theory, which enable all energy to be diffracted into a single Floquet order by engineering the scattering properties of the metaatom. The outstanding wavefront manipulation characteristics of metagratings offer a new design framework to passively focus the acoustic beam with scalable structures in an efficient way, providing an additional advantage of relaxed requirements on the fabrication resolution. Recent work at microwave frequencies revealed that metagratings can be combined with conventional gradient metasurfaces to improve the performance of metalenses with large numerical aperture, however such structures still require the fine discretization of a gradient metasurface in the central region of the lens (*35*). Nonetheless, this work suggested that the same concept of using a combination of different metagratings to converge beams might be valid also in acoustics.

Here, we demonstrate that an efficient metalens can be realized using adiabatically varying metagratings. We utilize the remarkable wavefront manipulation properties and anomalous reflection performance of metagratings to spatially control the beam diffraction along a metalens. Multilayered stepped structures, inspired to those used in Ref. (*33*), are utilized as metaatoms, due to their simple configuration. We establish a generalized semi-analytical model with a genetic algorithm to design a set of metagratings to locally control the excitation of a single Floquet order. Our metagrating approach, which combines multiple local metagratings with varying periodicity, can efficiently converge an ultrasonic wave to a focal spot. This versatile and scalable approach can overcome unfavorable energy dissipation of the metasurface-based lenses, to enable high efficiency acoustic focusing using passive means, while minimising the fabrication requirements and the effect of thermo-viscous losses.

**Results**

**A. Metagrating approach**

We propose a lens composed of multiple local metagratings to realize ultrasonic focusing, as illustrated in Fig. 1(a). Unlike conventional gradient metasurface-based lenses, we perform beam focusing by locally controlling the excitation of a single diffraction order, and without attempting to engineer a continuous phase distribution. This working principle can also be realized in a one-dimensional lens, as shown in Fig. 1(b). Assuming the lens is located at $y = 0$, in order to steer the acoustic beam towards a focal spot with a focal length of $y_f$, the local reflection angles $\theta_r(x)$ on the lens are calculated by the simple trigonometric relation

$$\tan \theta_r(x) = -x/y_f. \tag{1}$$

To perform this angle manipulation, we apply the metagrating concept, with locally varying diffraction angles. According to conventional grating theory, periodic structures diffract incident acoustic energy via all Floquet orders, in which the number of orders is determined by the periodicity of the grating. In our metagrating lens design, each metagrating is responsible for redirecting the normally incident acoustic wave towards a particular angle $\theta_n$ which is assigned as per Eq. (1). This angle is measured relative to the surface normal of the metagrating. The periodicity $L_n$ of the $n$-th metagrating is determined by the ratio of the acoustic wavelength of operation $\lambda$ to the reflection angle $\theta_n$ based on Bragg's condition

$$L_n = -\lambda/\sin \theta_n. \tag{2}$$

By combining different metagratings with varying local periods $L_n$, the directivity of beam diffraction along the metalens can be controlled, converging the acoustic beams to a focal spot. In this work, we consider the focal spot to be located at $x = 0$, which is aligned with the centre of the metalens as assumed in Eq. (1). We start by designing a half metalens (the region $x > 0$ in Fig. 1(b)), then subsequently mirror this structure about $x = 0$ and combine this with the original to obtain a full metalens. One unit cell is used in each metagrating. We use the adiabatically changing unit cell across the surface to smoothly vary the angle following Eq. (1).

### B. Design of local metagratings

Here, we aim to design each local metagrating to perform anomalous reflection in which all the acoustic energy is rerouted into a single Floquet mode (i.e., -1 for the right half-lens and +1 for the left half-lens) with the diffraction angle designed based on the condition expressed in Eq. (1). We previously showed that the scattering properties of the metaatom and the impedance mismatch between the incident wave and metagrating are the key design parameters in developing high efficiency metagratings for anomalous reflection (*34*). To properly manipulate the wavefront propagation towards a desired angle, the metaatom should be bianisotropic, which enables asymmetric scattering. By means of numerical and experimental analysis in Ref. (*33*), simple L-shaped structures have shown to be efficient means for beam steering in cases when only the (-1, 0 and +1) Floquet orders can propagate. When the desired reflection angle $|\theta_n|$ becomes smaller than or equal to 30°, higher Floquet orders appear in the system, such that a more advanced structure is required to simultaneously nullify numerous unwanted diffraction orders. In a metagrating-based lens, the centre of the lens must redirect the incident wave at small angles, therefore it requires a more complex structure to control higher order diffraction.

As such, we propose a multilayered stepped metaatom which is simple, yet powerful and scalable for rerouting acoustic waves into a single direction even for numerous higher Floquet orders being present. The metaatom is formed by $J$ layers of brick, as illustrated in Fig. 2(a) for $J = 3$. This multilayer structure provides the additional degrees of freedom to control the acoustic scattering properties of the local metagrating. The variation of thickness $h_j$ and width $l_j$ among different layers of brick enhances the bianisotopic properties of the metaatom, and subsequently support reflection with asymmetric amplitudes and allows anomalous reflection. This array of multilayered stepped structures is backed by a hard boundary such that high reflection efficiency is achieved. To predict the beam steering performance of the proposed metagratings, we develop a semi-analytical model with the full methodology outlined in Materials and Methods. This method involves the decomposition of the incident plane wave and the reflected field above the metagrating into Floquet modes, and the fields inside the multilayered air gap into waveguide modes. The reflection amplitudes into different Floquet orders can therefore be predicted by applying the pressure and velocity continuity conditions at the interfaces. The multilayered stepped structures are then optimized by a genetic algorithm as discussed in Materials and Methods.

## C. Lens design

We design a set of metagratings to realize high efficiency ultrasonic focusing of an incident plane wave to a focal spot at $y_f = 13\lambda$ at an operating frequency $f = 40$ kHz. According to the spatial distribution of the reflection angle and the Bragg's condition expressed as Eqs. (1) and (2), respectively, the first metagrating is designed to reroute the acoustic wave to a small angle $\theta_1 = -16°$. The design of this metagrating is the most challenging, since seven Floquet modes need to be taken into consideration (-3, -2, -1, 0, +1, +2 and +3). Figure 2(b) shows the analytical reflection spectrum of an infinite acoustic metagrating formed by an array of three-layer stepped metaatoms with $L_1 = 0.0311$ m, which is optimized to steer an acoustic beam towards $\theta_1 = -16°$. Its geometric parameters are listed in Table S1 in Supplementary Materials. Only (-3, -1 and +2) Floquet modes are shown in Fig. 2(b) since the reflection coefficients of the remaining unwanted diffraction orders are zero. The results reveal that the reflection coefficient for the -1 Floquet mode is 0.95, which indicates that almost all of the acoustic wave is reflected via the -1 Floquet mode at $f = 40$ kHz, while only a small part of it is scattered towards the -3 and +2 Floquet modes. Our semi-analytical results are verified by numerical simulations for the lossless case, with the resultant spectrum marked by crosses (×). The simulated scattered pressure field shown in the inset of Fig. 2(b) illustrates a highly planar reflected field, confirming the efficient anomalous reflection despite the numerous higher Floquet modes which can propagate. Using this semi-analytical approach, 26 metagratings are optimized for different reflection angles. The design of the full set of local metagratings and the optimized parameters of each metaatom are discussed in Supplementary Materials.

After designing these 26 metagratings, we combine them to form a half-metalens. Its focusing performance is investigated by two-dimensional full-wave simulations without thermo-viscous losses (see Materials and Methods). Figure 3(a) shows the resulting normalised intensity of the scattered field for the half-metalens, which is the intensity of the scattered pressure field $|p_{sca}|^2$ normalised by that of the incident pressure $|p_{inc}|^2$. With our metagrating approach, using only half a lens, focusing of a steered beam is already achieved. To quantify the size of the focal spot, we consider the scattered field profile along the line $y = 13.9\lambda$, shown by the white dotted line in Fig. 3(a). The resulting intensity is plotted in Fig. 3(c), showing that the focal spot is not vertically aligned with the left edge of the half-metalens (i.e., $x = 0$), but is shifted towards the left to $x = -1.25\lambda$. This shift occurs because the scattering phase centre of each metagrating does not coincide with the coordinate origin used when simulating the infinite metagrating.

The results of the half-metalens suggest that a tuning step is required before developing a full-metalens to precisely generate a fine and high intensity focal spot with low side lobes (see Supplementary Materials). To confirm the origin of the shifted focal spot, we establish a semi-analytical sound radiation model, where the multiple metagrating elements are simplified as an array of point sources. The semi-analytical result (dashed line) as indicated in Fig. 3(c) demonstrates that this method can successfully predict the shift of the focal spot $x_s = -1.25\lambda$.

To produce a full metalens with a focal point on its axis of symmetry $x = 0$, we first tune the half-metalens by shifting it horizontally by $x_s$, to compensate for the transverse shift of the focal spot. Combining this structure with its mirror half-metagrating leads to a full metalens with numerical aperture NA = 0.94, determined by the maximum reflection angle as NA = $\sin\theta_N$. Figures 3(b) and (d) show that a fine and high intensity focal spot is achieved by our full-metalens. The focusing performance of the metalens is quantitatively characterized in terms of the full width at half maximum (FWHM) and the maximum value of the normalized intensity of the scattered field $I_{max}$. The results illustrate that our proposed full-metalens enables acoustic focusing with maximum normalised intensity $I_{max} = 137.1$ at $f = 40$ kHz. Comparing with the half-metalens, the focal size FWHM reduces from $1.09\lambda$ to $0.364\lambda$. The small beam width of the focused main lobe is attributed to the interference of two steered beams generated by the two halves of the metalenses.

In the ultrasonic regime, thermo-viscous losses generally have an adverse impact on the acoustic focusing performance of metalenses. In gradient metasurface-based lenses, coiled space structures with long meander-line channels are commonly used to control the acoustic propagation path and, hence, to engineer the phase. High thermo-viscous losses are typically induced in such systems, due to the fact that the surface needs to be finely discretized to produce a continuously varying phase. The correspondingly narrow width of the internal geometries are often comparable to the viscous and thermal boundary layer thickness, which are ~9.5 µm at $f$ = 40 kHz (*36*), resulting in dissipation of the acoustic energy and degradation of the focusing performance (*25*, *37*). Our metaatoms are comparable with the wavelength, which largely overcomes this limitation. We examine the effect of thermo-viscous losses on our proposed metagrating lens by a finite element model, with the resultant spectrum shown in Fig. 3(d). Comparing the lossless and lossy cases, only a small reduction of 3% in the maximum intensity is found. This demonstrates that our proposed metagrating approach with simple stepped structures can overcome the unfavorable energy dissipation and fabrication complexity of gradient metasurface-based lenses for ultrasonic focusing.

### D. Experimental verification

To validate our metagrating-based design principle, the focusing performance of our ultrasonic metalens is experimentally demonstrated and compared with numerical predictions (Fig. 4). Our experiments are conducted inside an anechoic chamber with the setup shown in Fig. 4(a). Although the theoretical and numerical models assume focusing along a single axis, we demonstrate here that they accurately predict the performance of a cylindrically symmetric lens, which focuses the fields along two axes. Such focusing is most relevant for applications, and it leads to a strong focal spot which is easiest to measure experimentally. The metalens is fabricated by 3D printing with Polylactic Acid (PLA). As the 3D printer has a restricted build volume, a metalens with smaller numerical aperture NA = 0.76 is considered. The evolution of the two-dimensional metalens can be realised by simply revolving a one-dimensional design of the half-metalens about its $y$-axis as indicated in Fig. 1(b). A single ultrasonic transducer located at ($7\lambda$, $105\lambda$) is used to generate the incident wave at $f = 40$ kHz. The transducer is shifted from the $y$-axis of the metalens to suppress multiple reflections between the lens and transducer. To better compare the numerical and experimental results, simulations are conducted using a spherical sound source at the coordinates of the transducer.

Figures 4(b) and (c) show the simulated and the measured distribution of the normalized intensity of the scattered field, respectively. Compared with the simulated results for plane wave excitation, a shifted and tilted focal spot of elliptical shape at around (-$1.3\lambda$, $16.8\lambda$) is found both numerically and experimentally due to the off-axis source location (see Supplementary Materials). Significant intensity enhancement is obtained at the focal spot, which is nearly 22.9 times of the incident one. Since the simulations are calculated in a 2D domain with a 1D metalens, while the measurements are performed with a 2D metalens, the results are normalized by their maximum normalised intensity $I_{max}$. The profiles of normalized intensity on the focal plane along the transverse and axial directions are illustrated in Figs. 4(d) and (e), respectively, showing that our experimental results are in good agreement with the numerical prediction. The measured FWHM is $0.6\lambda$, which is 9% lower than the numerical result.

**Discussion**

We proposed a straightforward design principle for efficient ultrasonic beam focusing by using adiabatically varying metagratings to locally control the excitation of a single Floquet order. The convergence of acoustic energy to a focal spot with designated focal length is achieved by spatially engineering the directivity of diffracted beams along the metalens. We apply the concept of metagratings to satisfy this spatially varying profile in accordance with grating theory. To effectively diffract the acoustic wave via a particular Floquet mode, which is aligned with the desired local angle, the scattering properties of the metagratings are engineered, such that reflections towards unwanted directions are minimised. We propose a multilayered stepped structure as the metaatom, where its multiple layers offer additional degrees of freedom to effectively control the asymmetric scattering properties. This enables the metaatom to achieve high efficiency anomalous reflection, i.e., $R_{-1} > 0.95$, especially for elements near the center of the lens which must control numerous higher order Floquet modes (up to 7 modes in this study). A semi-analytical model is incorporated into a genetic algorithm to optimize the designs of the metaatoms for a variety of reflection angles, in which their beam steering performance is verified numerically ranging from -16° to -78°. At the operating frequency of $f = 40$ kHz, a set of metagratings is designed. Compared to the space-coiling structures commonly used as gradient metasurface, our metagratings promise to be easier scaled down for applications at higher frequencies, since narrow and complex internal geometries are avoided. Meanwhile, the absence of the long labyrinthine meander also minimizes thermo-viscous losses, such that the wavefront manipulation and focusing performance can be maintained when scaling the structure over a wide range of wavelengths.

We developed the metalens by combining the optimized metagratings to satisfy the spatial reflection angle profile for acoustic focusing. A fine-tuning step is required to minimize the side lobes which are induced due to the discontinuous change of the reflection angles between the metagratings. The required tuning distance $x_s$ can be estimated using our semi-analytical sound radiation model, where the multiple metagrating elements are simplified as an irregular array of point sources. The numerical results demonstrate that the developed metagrating based lens can perform intensive beam focusing with equivalent performance to a gradient metasurface-based lens at ultrasonic frequencies, while less dense discretization is required. Its promising acoustic focusing performance has also been experimentally demonstrated. This metagrating approach can be adapted to perform acoustic focusing for different operating frequencies and focal lengths, which may facilitate the further development of a wide range of application devices including acoustic levitators and tweezers, diagnosis systems, or technology used in ultrasound therapy.

**Materials and Methods**
**Semi-analytical model for metagrating design**
When a normally incident plane wave impinges on a periodic structure of periodicity $L$, the acoustic wave is diffracted via different Floquet modes. The total acoustic field above the metagrating $p_a$ can be expressed as the sum of incident and reflected waves

$$p_\mathrm{a}(x,y) = p_0 e^{ik_0 y} + \sum_{m=0}^{+\infty} B_m e^{-iG_m x - i\beta_m y}, \qquad (3)$$

where $p_0$ is the amplitude of incident wave, $k_0$ is the wavenumber in free space, $B_m$ denotes the amplitude of reflected pressure of $m$-th Floquet mode, $G_m$ and $\beta_m$ are the wavenumbers of $m$-th mode scattering which can be written as $G_m = 2\pi m/L$ and $\beta_m = \sqrt{k_0^2 - G_m^2}$, respectively.

Considering the stepped structure as illustrated in Fig. 2(a), the air gap between each metaatom can be divided into multiple sub-layers. The depth and width of the *j*-th sub-layer of the air gap are denoted by $h_j$ and $L - l_j$ (i.e., j = 1, 2, ..., *J*), respectively, where the acoustic pressure inside can be expressed as a superposition of waveguide modes

$$p_j(x,y) = \sum_{q=0}^{\infty} \cos \alpha_{q,j}(x - l_j) \left[ C_{qj}^+ e^{i\gamma_{qj}y} + C_{qj}^- e^{-i\gamma_{qj}y} \right], \quad (4)$$

where $C_{qj}^+$ and $C_{qj}^-$ are the amplitudes of the *q*-th order of waveguide mode in the -*y* and the +*y* directions, respectively, $\alpha_{q,j}$ and $\gamma_{qj}$ are the wavenumbers of the *q*-th waveguide mode which can be written as $\alpha_{q,j} = q\pi/(L - l_j)$ and $\gamma_{qj} = \sqrt{k_0^2 - \alpha_{q,j}^2}$, respectively.

The normal particle velocity along *y*-direction above the metagrating and inside the air gap can be obtained by

$$v = -\frac{1}{i\omega\rho_0}\frac{\partial p}{\partial y}, \quad (5)$$

where $\omega$ denotes the angular frequency and $\rho_0$ is the density of air.

By applying the pressure and velocity continuity boundary conditions at the opening of air gap and the interfaces between each air sub-layer, we obtain a series of equations, and the reflection coefficients of different Floquet orders can then be solved by using the orthogonality relationship of the Floquet and waveguide modes, as described in Ref. (*31*). In this study, 19 waveguide modes are considered to analytically calculate the reflection coefficient $R_m$ of the local metagratings

$$R_m = \frac{|B_m|^2 \cos \theta_m}{|p_0|^2}, \quad (6)$$

where $\theta_m$ is the reflection angle of *m*-th Floquet mode.

The genetic algorithm with continuous variables is applied to the above semi-analytical model to define the optimized $h_j$ and $l_j$ (j = 1, 2, ..., *J*) of the metagratings. In this study, the role of the metagratings is to reflect the acoustic wave towards a single direction aligned with the -1 Floquet order. As such, in the optimization algorithm, we set the target function as $F = \max(R_{-1})$ to search for the optimal geometric parameters of the metagratings for achieving high efficiency anomalous reflection.

**Numerical simulations**

To investigate the focusing performance and validate the semi-analytical model, 2D full-wave numerical simulations are performed with COMSOL Multiphysics 5.4. For the lossless case, the simulations are performed with the Pressure Acoustics Module, while the Multiphysics

Thermoviscous Acoustics Module is also included for the case with losses. The acoustic boundaries of the metaatoms are treated as sound hard boundaries. The mechanical boundary is set to be no slip and the thermal boundary is set to be isothermal for the lossy simulations. Perfectly matched layers are added to the outer boundaries of simulation domains to avoid reflections. The maximum element size is set to $\lambda/10$.

**Acoustic field measurements**

To experimentally demonstrate the acoustic focusing performance of the metalens, acoustic field measurements were conducted in an anechoic chamber with the experimental setup shown in Fig. 4(a). The incident wave is generated by an ultrasonic transducer at frequency $f = 40$ kHz by using a Teensy 3.2 micro-controller development board. The field measurements was obtained using a 1/4 in. microphone (B&K type 4135) connected to a microphone power supply (B&K type 2807), which was moved in two axes driven by a translation stage.

The 2D metalens was 3D printed via fused deposition modelling with Polylactic Acid (PLA). One quarter of the lens was printed each time due to the building volume limitation of the 3D printer. These four parts of the metalens were assembled by gluing onto a cardboard backing. The metalens was mounted on brackets, and placed at a distance of 0.9m away from the transducer. The brackets was surrounded by absorbing foam to prevent unwanted reflection.

**Acknowledgments**

**Funding:** DP and SO acknowledge funds of the Australian Research Council Discovery Project DP200101708. LQ, YP and AA have been supported by the National Science Foundation and the Simons Foundation.

**Author contributions:** Y.K.C.: Conceptualization, Methodology, Software, Validation, Investigation, Data Curation, Writing – Original Draft, Visualization. L.Q.:

Methodology, Writing – Review & Editing. Y.P.: Methodology, Writing – Review & Editing. S.S.: Methodology, Writing – Review & Editing. S.O.: Conceptualization, Resources, Writing – Review & Editing, Project Administration, Funding acquisition. A.A.: Conceptualization, Writing – Review & Editing, Supervision, Project Administration, Funding Acquisition. D.P.: Conceptualization, Investigation, Resources, Writing – Review & Editing, Supervision, Project Administration, Funding Acquisition.

**Competing interests:** The authors declare that they have no competing interests.

**Data and materials availability:** All data needed to evaluate the conclusions in the paper are present in the paper and/or the Supplementary Materials. Additional data related to this paper may be requested from the corresponding authors.

**Figures and Tables**

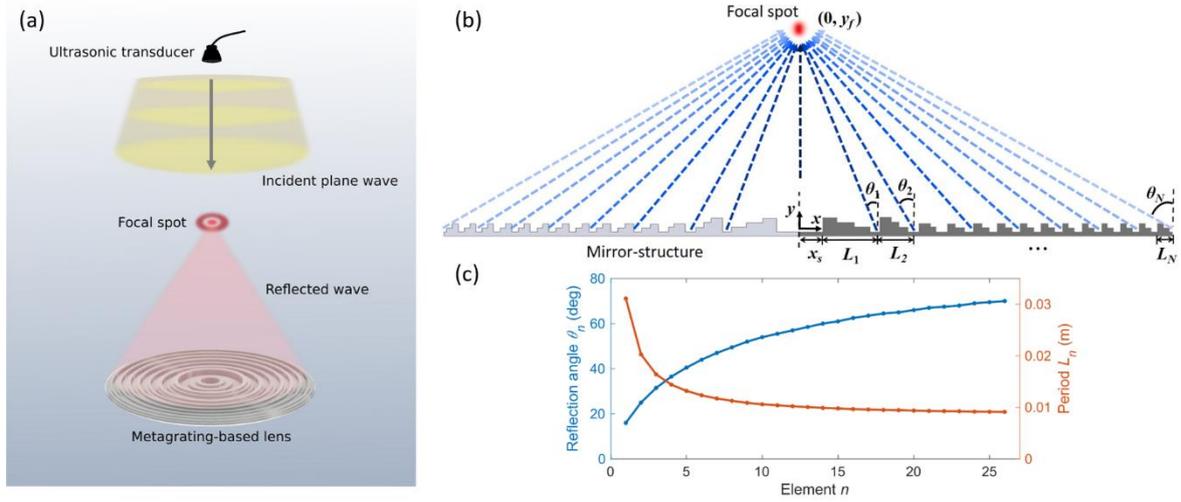

**Fig. 1. Metagrating for acoustic focusing.** (a) Schematic of the metagrating-based reflective lens for acoustic focusing. (b) Illustration of the design principle of the proposed metagrating approach, which can converge the acoustic energy to a focal spot by spatially control the reflection angles. (c) Corresponding reflection angle and period of each metagrating element to perform focusing for $y_f = 13\lambda$ at $f = 40$kHz.

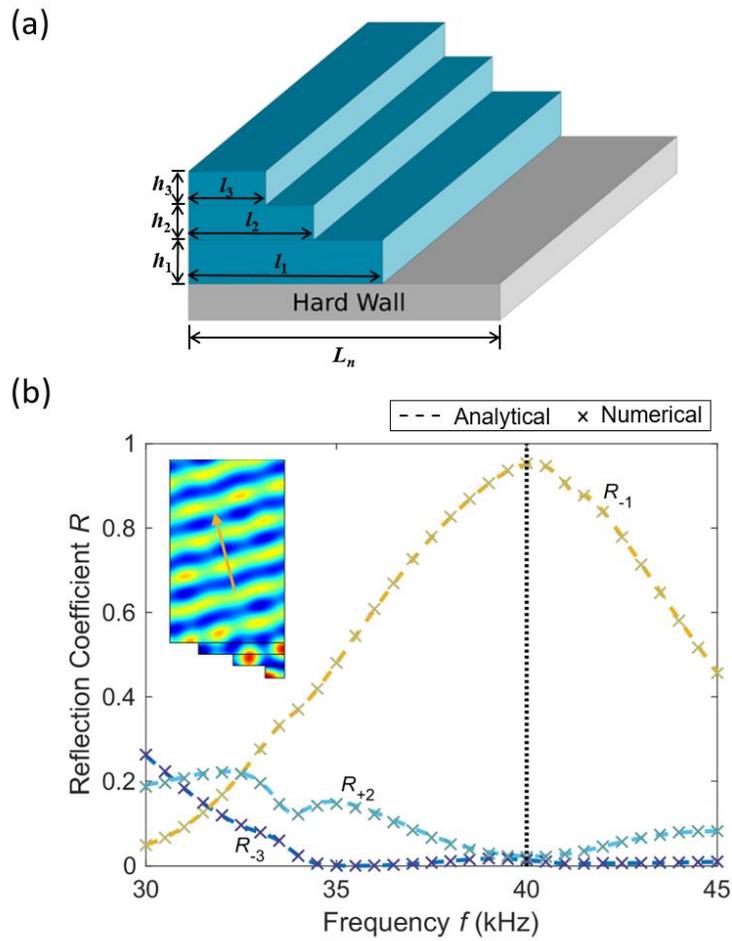

**Fig. 2. Design of Metaatoms.** (a) Configuration of the proposed metagrating composed of multilayered stepped meta-atoms. (b) Reflection coefficient of our designed metagrating with seven Floquet modes. Inset shows the real part of reflected pressure fields (-1 Floquet mode) at $f$ = 40kHz.

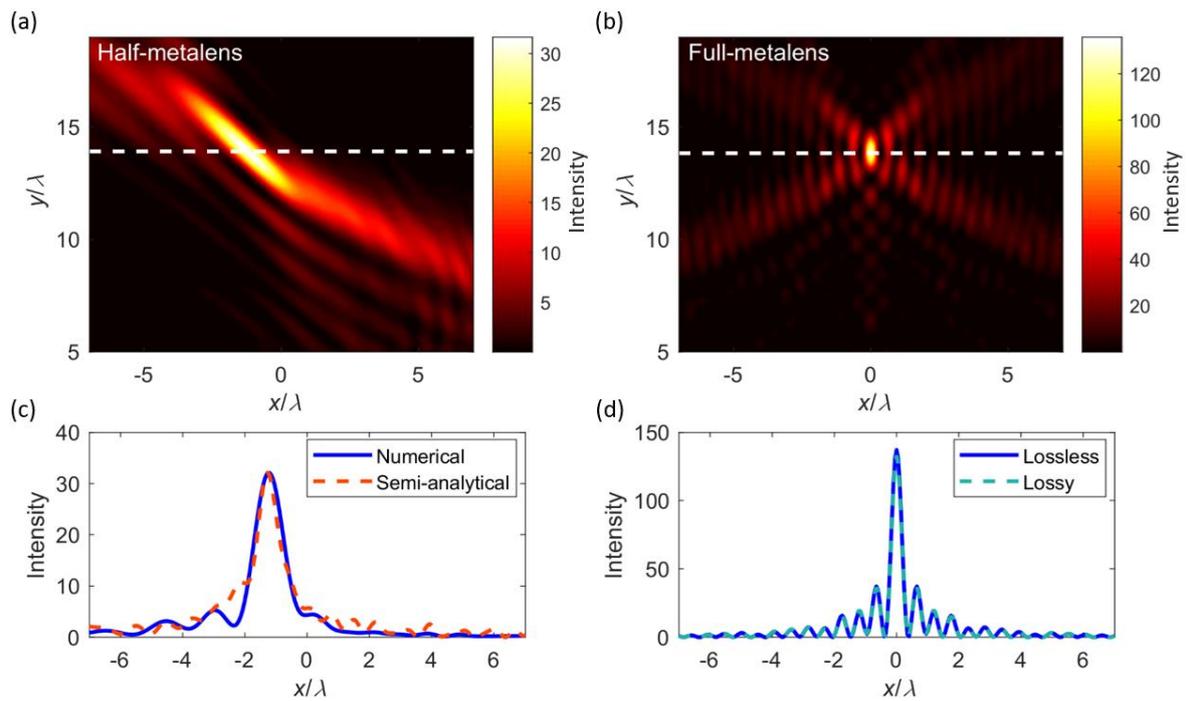

**Fig. 3. Modelling beam focusing.** (a), (b) Numerical simulations of the two-dimensional intensity normalized to the intensity of the incident plane wave distributions by using 1D half and full metalenses, respectively. (c) Numerically simulated (solid line) and semi-analytically predicted (dashed line) normalized intensity distributions along the focal plane $y = 13.9\lambda$ as indicated by the white dashed lines in (a) for the half-metalens. (d) Comparison of the numerical simulated normalized intensity distributions between the lossless (solid line) and lossy (dashed line) cases along the focal plane $y = 13.8\lambda$ as indicated by the white dashed lines in (b) for the full-metalens. The simulated FWHMs of the half and full size cases are $1.09\lambda$ and $0.364\lambda$, respectively.

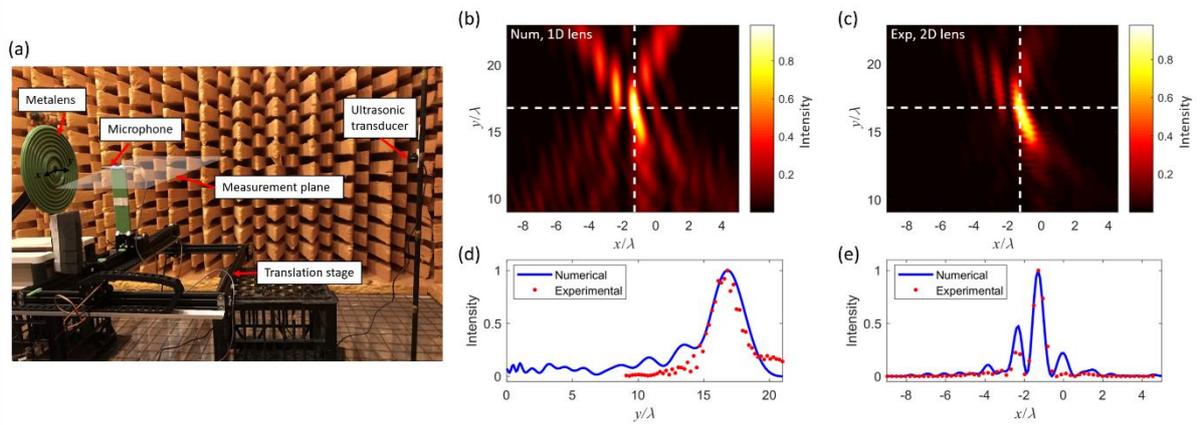

**Fig. 4. Experimental demonstration of ultrasonic focusing with a metagrating lens.** (a) Snapshot of the experimental setup. (b), (c) Numerically simulated and experimentally measured normalized reflected intensity in the measurement plane indicated in (a). (d), (e) Numerically predicted (solid line) and experimentally measured (·) normalized reflected intensity distributions in cross sections of the focal plane along y and x directions, respectively. The white dashed lines in (b) and (c) indicate where the results in (d) and (e) are extracted. The 1D numerical results and 2D experimental results are normalized to their maximum intensity to enable comparison.